\documentclass[11pt,a4paper]{article}
\usepackage{jheppub_kim}
\usepackage{epsfig}
\usepackage{amsmath}
\usepackage{graphicx}
\usepackage{graphics}
\usepackage[scriptsize,nooneline,hang]{caption}
\usepackage[hang,nooneline,scriptsize]{subfigure}

\begin{document}

\title{  Geodesic Motion in the Spacetime of a SU(2)-Colored (A)dS Black Hole in Conformal Gravity }

\author[a]{Bahareh Hoseini,}
\author[*a]{ Reza Saffari,}
\affiliation[a]{Department of Physics, University of Guilan, 41335-1914, Rasht, Iran}
\author[b]{ Saheb Soroushfar}
\affiliation[b]{Faculty of Technology and Mining, Yasouj University, Choram 75761-59836, Iran}

\emailAdd{rsk@guilan.ac.ir}
\emailAdd{soroush@yu.ac.ir}

\abstract{In this paper, we study the geodesic motion in the spacetime of a SU(2)-colored (A)dS black hole in conformal gravity, and also we investigate spacetime features, such as light spheres and horizons. Moreover, we derive the analytical solutions for the equation of motion of test particles and light rays using Weierstrass elliptic and Kleinian $\sigma$ functions. Depending on the particle energy levels and angular momentums, we classify the solutions of the geodesic equations. Furthermore, several examples of possible types of orbits illustrate the results.}

\maketitle

\section{Introduction}
One of the gravity theories with higher-order derivatives in action is Conformal (Weyl) Gravity (CG). This theory can solve the dark matter and dark energy problems and involve quantum gravity \cite{Lu:2011zk,Maldacena:2011mk,Mannheim:1988dj}. Maxwell or general non-abelian Yang-Mills fields are invariant under conformal transformations. Using these gauge fields and CG, Fan and Lu presented some solutions for a black hole with SU(2) Yang-Mills charges. \cite{Fan:2014ixa}. On the other hand, one method to investigate gravitational fields of objects like black holes is studying the motion of test particles around them. For this reason, studying the motion of objects in the spacetime of a SU(2)-colored (A)dS black hole in CG helps us to study this spacetime. Weierstrassian elliptic functions are helpful to study the analytic solution of motion. Hagihara, in 1931 demonstrated the analytical solution for Schwarzschild spacetime in terms of elliptic functions for the first time. \cite{Y. Hagihara}. Also, this method has studied for Kerr de Sitter spacetime \cite{C. Lammerzahl}, Schwarzschild de Sitter in 4-dimensional spacetime \cite{E. Hackmann}, Reissner-Nordstrom spacetime, and Reissner-Nordstrom-(anti) de Sitter spacetime \cite{V. Z. Enolski}. Besides, the geodesic equations are solved analytically in some spacetime such as BTZ, dilaton black holes, the spacetime of black holes in CG, $f(R)$ gravity, Rastall theory, and black string \cite{Soroushfar:2015dfz,Soroushfar:2016yea,Hoseini:2016nzw,Hoseini:2016tvu,Soroushfar:2015wqa,Soroushfar:2016esy,Soroushfar:2018yzw,Rezvanjou:2017hox}.

This paper is interested in solving the geodesic equations of a SU(2)-colored (A)dS black hole, while this type of blake hole is rooted in CG \cite{Fan:2014ixa}. It is necessary to mention that there are some different descriptions of the applicability of geodesic solutions in CG. While some investigations predicate that only null geodesics are physically meaningful in CG \cite{Brihaye:2009ef,Barabash:1999bj,Wood:2001ve}, other discussions state that timelike geodesic is also applicable, such as free fall of elementary particles bound packets or geons gravitational energy packets, which Wheeler suggests \cite{Ohanian:2015wva}. The possibility of choosing the gauge, which makes this theory attractive for all geodesics, is studied in Ref. \cite{Edery:2001at}.

Furthermore, derivation of the analytical solutions for the equation of motion is planned here in terms of Weierstrass elliptic and derivatives of Kleinian sigma functions in the spacetime of a SU(2)-colored (A)dS black hole. The region of test particle motion is studied using significant potential and real zeros of a polynomial. The organization of this paper is as follows: Sec.~\ref{FE}, is an introduction to the Lagrangian and field equation of a SU(2)-colored (A)dS black hole spacetime in CG. Sec.~\ref{GE}, derives the geodesic equations, Sec.~\ref{FSTG}, studies features of the spacetime geometry, Sec.~\ref{AS}, gives analytical solutions, Sec.~\ref{O}, shows the possible orbit illustration, and the conclusion attains in Sec.~\ref{C}.

\section{SU(2)-colored (A)dS black hole in conformal gravity with only SU(2) Yang-Mills charges}\label{FE}
Conformal invariance is preserved for the Maxwell or non-abelian YangMills fields because these fields are additionally conformal. As mentioned in Ref.~\cite{Fan:2014ixa}, using this Lagrangian
\begin{equation}\label{EF}
\mathcal{L}=\alpha
\sqrt{-g}(\frac{1}{2}C_{\mu\nu\rho\sigma}C^{\mu\nu\rho\sigma}
-\frac{1}{2g_s^2}F^a_{\mu\alpha}F^{a\mu\alpha}-\frac{1}{2e^2}\mathcal{F}_{\mu\nu}\mathcal{F}^{\mu\nu}),
\end{equation}
in which $F^a$ and $\mathcal{F}$ are the field strengths of the
$SU(2)$ Yang-Mills and $U(1)$ gauge fields
($F^a_{\mu\nu}=\partial_\mu A^a_{\nu}-\partial_\nu
A^a_{\mu}+\epsilon^{abc}A^b_{\mu}A^c_{\nu},\quad
\mathcal{F}_{\mu\nu}=\partial_\mu \mathcal{A}_{\nu}-\partial_\nu
\mathcal{A}_{\mu}$), and $C^{\mu\nu\rho\sigma}$ is the Weyl tensor,
and one can achieve this metric
\begin{equation}\label{metric1}
ds^2=-f(r)dt^2+\frac{dr^2}{f(r)}+r^2(d\theta^2 +\sin^2 \theta
d\phi^2),
\end{equation}
with
\begin{equation}\label{f1}
f=a_0r^2(1+\frac{b_1}{(b_0-1)r})^2(1+\frac{b_1}{(b_0+1)r})^2.
\end{equation}
For simplification, Eq.~(\ref{f1}) can be rewritten as
\begin{equation}\label{fr}
f(r)=a_0r^2+a_0cr+a_0(c^2+2d)+2a_0dc\frac{1}{r}+a_0d^2\frac{1}{r^2},
\end{equation}
where
\begin{equation}\label{cd}
\frac{b_1}{(b_0-1)}+\frac{b_1}{(b_0+1)}=c,\quad \frac{b_1}{(b_0-1)}
\frac{b_1}{(b_0+1)}=d.
\end{equation}
The above metric can be asymptotic to AdS spacetime for $\Lambda = -3a_0 <0$, \cite{Fan:2014ixa}. This solution has a curvature singularity at $r = 0$. Solving $f(r_0)=0$, locates one or more event horizons. So the roots of function $f$ are
\begin{equation}
r_1=-\frac{b_1}{b_0-1},\qquad r_2=-\frac{b_1}{b_0+1}.
\end{equation}
The following states can be considered for $b_1<0$, \cite{Fan:2014ixa}:
\begin{enumerate}
\item  If $b_0>1$, then $r_1$ and $r_2$ are positive and $r_1>r_2$. In this case, there is a single black hole, and the spacetime outside its horizon is $r\geq r_1$.
\item  If $-1<b_0<1$, then $r_1<0$ and $r_2>0$, there are two black holes with two event horizon $r=r_1$ and $r=r_2$, and they are asymptotic to AdS as $r=-\infty$ and $r=+\infty$, respectively.
\item If $b_0<-1$, then $r_2<r_1<0$, and there is a single black hole. The event horizon for this black hole is $r=r_2$, and the spacetime outside of this horizon determine by $-\infty<r<r_2$.
\end{enumerate}

\section{The geodesic equation}\label{GE}
For the metric of Eq.~(\ref{metric1}), there are two conserved
quantities; energy and angular momentum as
\begin{equation}\label{E}
E=g_{tt}\frac{dt}{ds}=\frac{dt}{ds}(a_0r^2+2a_0cr+a_0(c^2+2d)+2a_0dc\frac{1}{r}+a_0d^2\frac{1}{r^2}),
\end{equation}
\begin{equation}\label{L}
L=g_{\varphi\varphi}\frac{d\varphi}{ds}=r^2\frac{d\varphi}{ds}.
\end{equation}
The Lagrangian $\mathfrak{L}$ for a test particle in this spacetime
is
\begin{equation}\label{lag}
\mathfrak{L}=\frac{1}{2}g_{\mu\nu}\frac{dx^{\mu}}{ds}\frac{dx^{\nu}}{ds}=\frac{1}{2}\epsilon.
\end{equation}
Using Eqs.~(\ref{E}--\ref{lag}), geodesic equations convert to the
following differential equation
\begin{equation}\label{rtu}
(\frac{dr}{d\tau})^2=E^2-f(r)(\epsilon +\frac{L^2}{r^2}).
\end{equation}
The two differential equations for $r$ as a function of $\phi$ and as a function of $t$ are
\begin{equation}\label{rphi}
(\frac{dr}{d\phi})^2=\frac{r^4}{L^2}(E^2-f(r)(\epsilon
+\frac{L^2}{r^2}))=R(r).
\end{equation}
\begin{equation}\label{rt}
(\frac{dr}{dt})^2=\frac{f(r)}{E^2}(E^2-f(r)(\epsilon
+\frac{L^2}{r^2})).
\end{equation}
Eqs.~(\ref{rtu})-(\ref{rt}) give a complete description of the
dynamics. Eq.~(\ref{rtu}) suggests the introduction of an effective
potential
\begin{equation}\label{veff}
V_{eff}=(a_0r^2+2a_0cr+a_0(c^2+2d)+2a_0dc\frac{1}{r}+a_0d^2\frac{1}{r^2})(\epsilon
+\frac{L^2}{r^2}).
\end{equation}

\section{Features of the spacetime geometry }\label{FSTG}
In this section, features of spacetime such as light spheres and horizons are discussed. For this purpose, effective potential and equation of motion for photon orbits around a compact gravitational source are considered.
With the dimensionless quantities,
\begin{equation}\label{DP}
\tilde{r}=\frac{r}{M}, \qquad
\tilde{c}=\frac{c}{M}, \qquad
\tilde{d}=\frac{d}{M^{2}}, \qquad
\tilde{a_0}=M^2 a_0, \qquad
\mathcal{L}=\frac{M^{2}}{L^{2}},
\end{equation}
Eq.~(\ref{veff}) can be written as
\begin{equation}\label{VD}
V_{eff}=\bigg(\tilde{a_0}\tilde{r}^2+2\tilde{a_0}\tilde{c}\tilde{r}+\tilde{a_0}(\tilde{c}^2+2\tilde{d})+2\tilde{a_0}\tilde{d}\tilde{c}\frac{1}{\tilde{r}}+\tilde{a_0}\tilde{d}^2\frac{1}{\tilde{r}^2}\bigg)\bigg(
\frac{1}{\mathcal{L}\tilde{r}^2}\bigg).
\end{equation}

\subsection{Light spheres}
In order to study the feature of spacetime, such as a light sphere, the effective potential derivative is examined. The light spheres lie at the stationary points of the effective
potential \cite{Turner:2020gxo},
\begin{equation}\label{dV}
\dfrac{dV_{eff}}{d\tilde{r}}=-2\,{\frac {\tilde{a_{0}} \left( {\tilde{c}}^{2}{\tilde{r}}^{2}+\tilde{c}{\tilde{r}}^{3}+3\,\tilde{d}\tilde{c}\tilde{r}+2\,\tilde{d}{\tilde{r}}^{2}+2\,{\tilde{d}}
		^{2} \right) }{\mathcal{L}{\tilde{r}}^{5}}}=0,
\end{equation}
so,
\begin{equation}\label{dV}
\tilde{r}_{1}=-\tilde{c}/2-1/2\,\sqrt {{\tilde{c}}^{2}-4\,\tilde{d}}, \qquad
\tilde{r}_{2}=-2\,{\frac {\tilde{d}}{\tilde{c}}}, \qquad
\tilde{r}_{3}=-\tilde{c}/2+1/2\,\sqrt {{\tilde{c}}^{2}-4\,\tilde{d}},
\end{equation}
are the radii of three light spheres with positions independent of $ \tilde{a_{0}}$ where $\tilde{r}_{1}<\tilde{r}_{2}< \tilde{r}_{3}$. The effective potential informs the stability of the light spheres, so the minima at $\tilde{r}_{1}=-\tilde{c}/2-1/2\,\sqrt {{\tilde{c}}^{2}-4\,\tilde{d}}$ and $\tilde{r}_{3}=-\tilde{c}/2+1/2\,\sqrt {{\tilde{c}}^{2}-4\,\tilde{d}}$ correspond to a stable light sphere, while the maximum in $V_{eff}$ at $\tilde{r}_{2}=-2\,{\frac{\tilde{d}}{\tilde{c}}}$ corresponds to an unstable light sphere.

\subsection{Horizons}
Solutions of the metric (\ref{fr}) in its dimensionless form represent the horizons by setting the condition $f(\tilde{r})=0$. So,
\begin{equation}
\tilde{a}{\tilde{r}}^{2}+2\,\tilde{a}\tilde{c}\tilde{r}+\tilde{a} \left( {\tilde{c}}^{2}+2\,\tilde{d} \right) +2\,{\frac {\tilde{a}\tilde{d}\tilde{c}}{\tilde{r}}}+{
	\frac {\tilde{a}{\tilde{d}}^{2}}{{\tilde{r}}^{2}}}=0,
\end{equation}
has two real solutions, such as
\begin{equation}\label{dV}
\tilde{r}_{h1}=-\tilde{c}/2-1/2\,\sqrt {{\tilde{c}}^{2}-4\,\tilde{d}}, \qquad
\tilde{r}_{h2}=-\tilde{c}/2+1/2\,\sqrt {{\tilde{c}}^{2}-4\,\tilde{d}},
\end{equation}
which are the horizons' locations for this spacetime. Moreover, from the effective potential (\ref{VD}), one can say that horizons occur where $ V_{eff}=0 $. In the following, we use ‘in’ and ‘out’ for ‘interior’ and ‘exterior’ horizons or stable light spheres (see Tab. \ref{tab:HS}).

\begin{figure}[h]
	\centering
	\subfigure[$ \tilde{c}>0 $ , $ \tilde{d}\geq 0 $]{
		\includegraphics[width=0.45\textwidth]{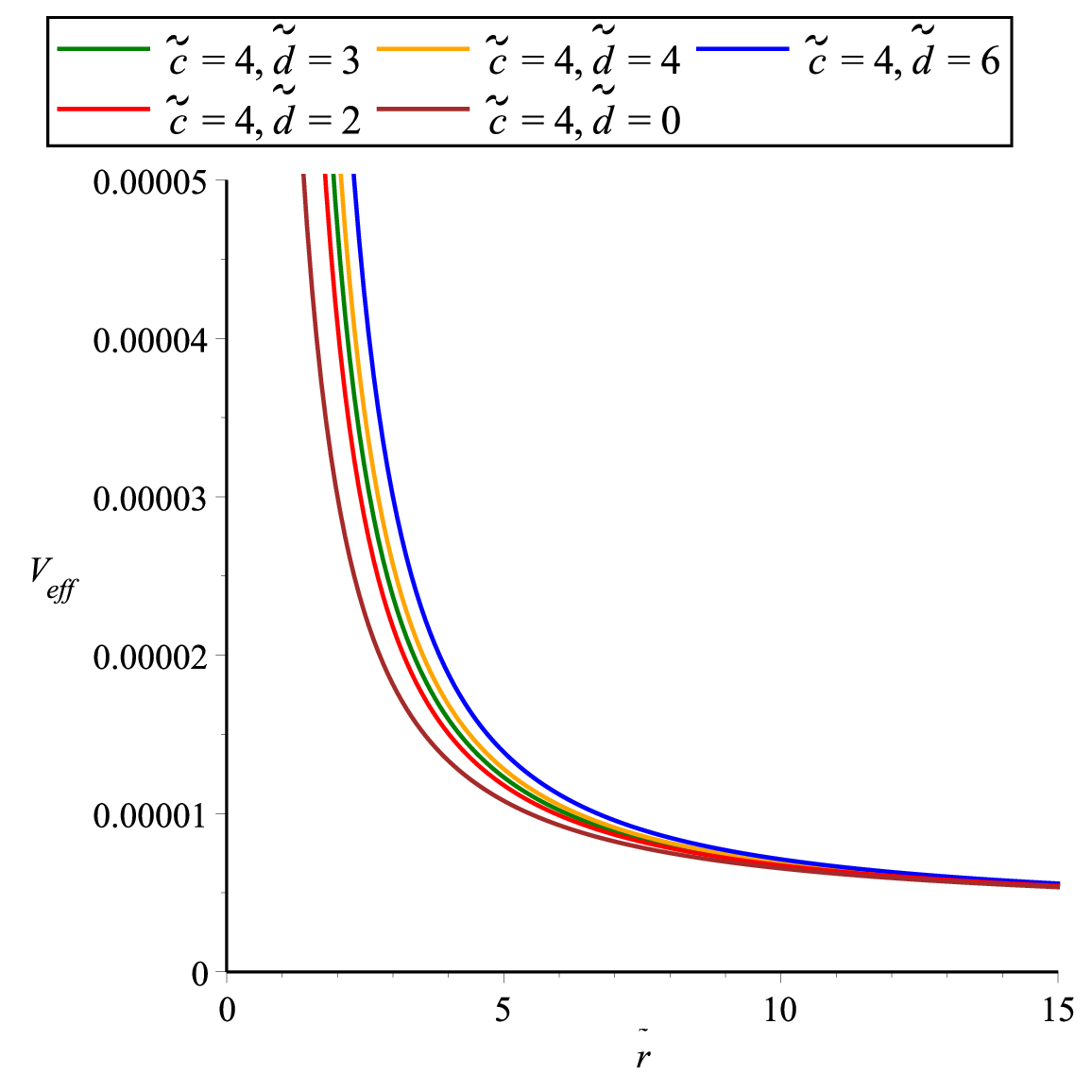}
	}
	\subfigure[$ \tilde{c}>0 $ , $ \tilde{d}< 0 $]{
		\includegraphics[width=0.45\textwidth]{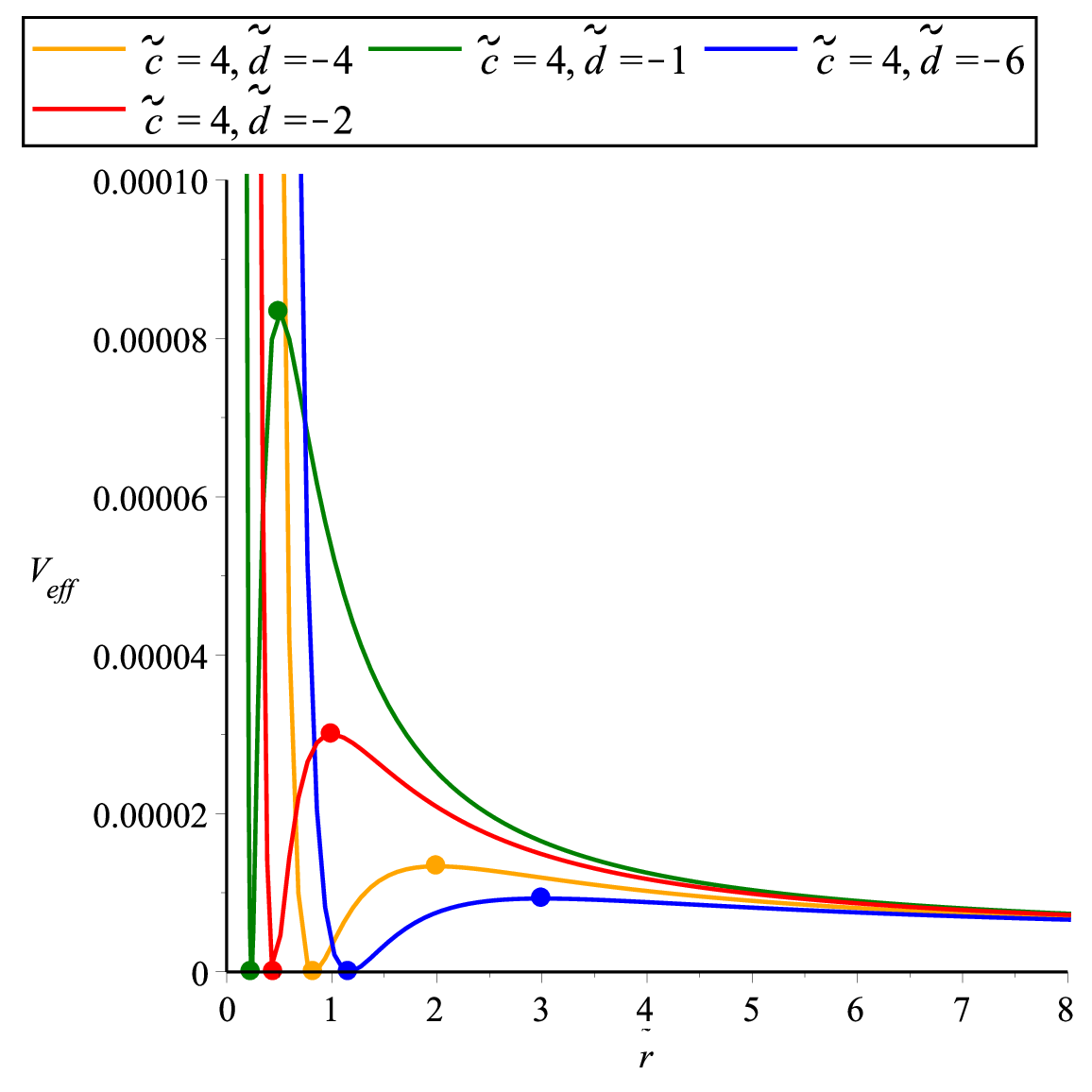}
	}
	\subfigure[$ \tilde{c}<0 $ , $ \tilde{d}\geq 0 $]{
		\includegraphics[width=0.45\textwidth]{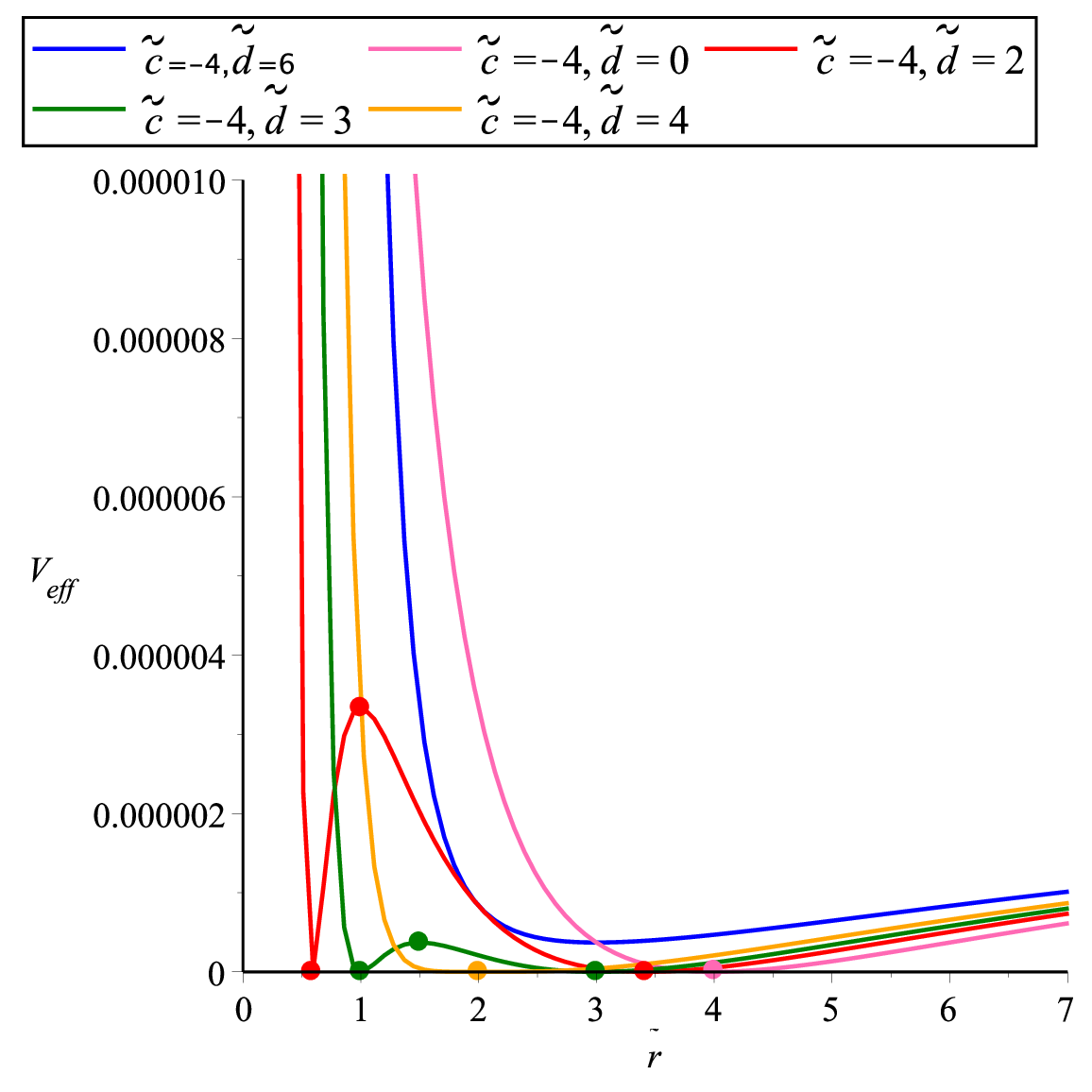}
	}
	\subfigure[$ \tilde{c}<0 $ , $ \tilde{d}< 0 $]{
		\includegraphics[width=0.45\textwidth]{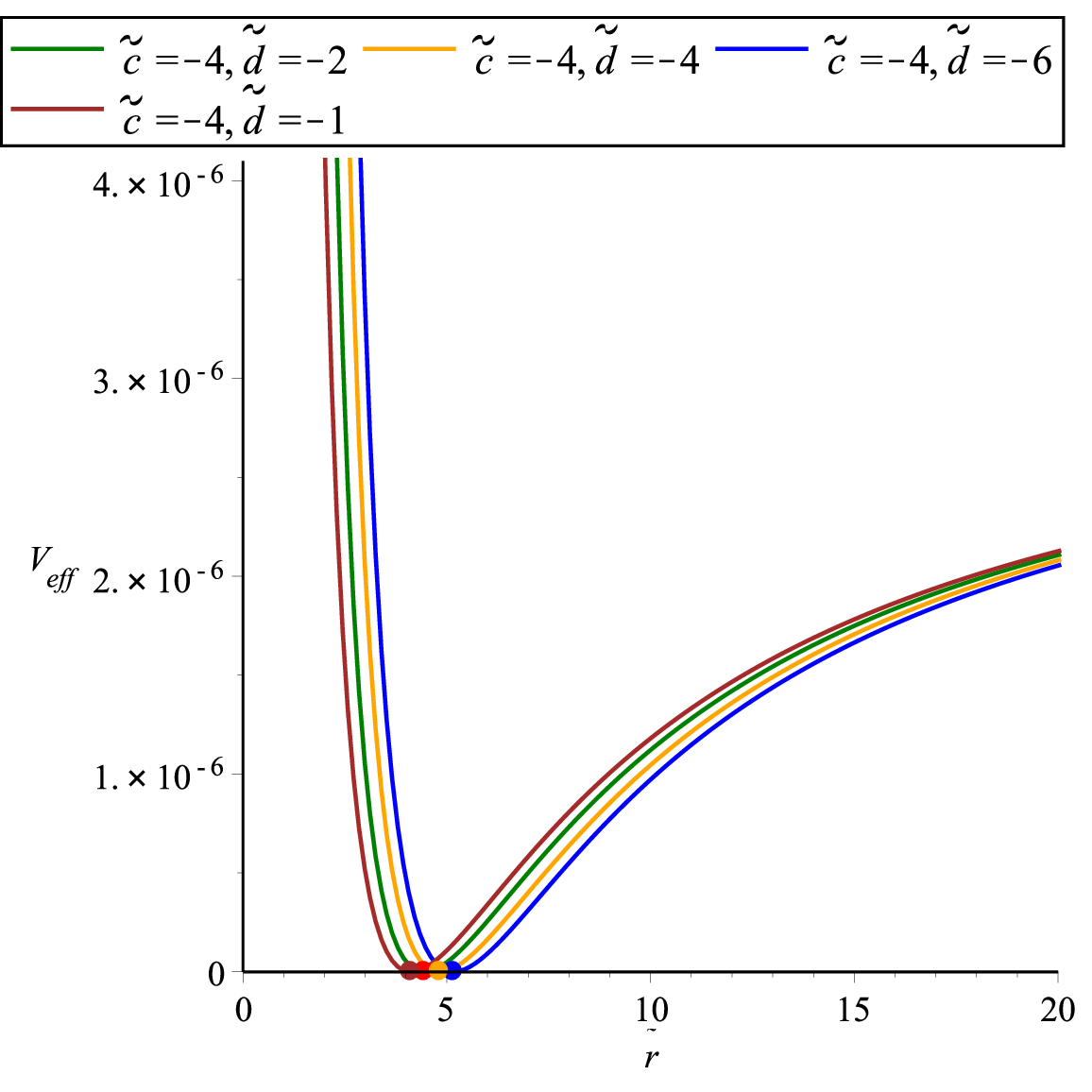}
	}
	\caption{Effective potentials of photon orbits as a function of radius ($ \tilde{r} $) for a range of the parameters
$ \tilde{c}$ and $ \tilde{d} $. The positions of light spheres are marked by dots.}
	\label{FVD}
\end{figure}

\begin{figure}[h]
	\centering
	\subfigure[ $ d>0 $]{
		\includegraphics[width=0.4\textwidth]{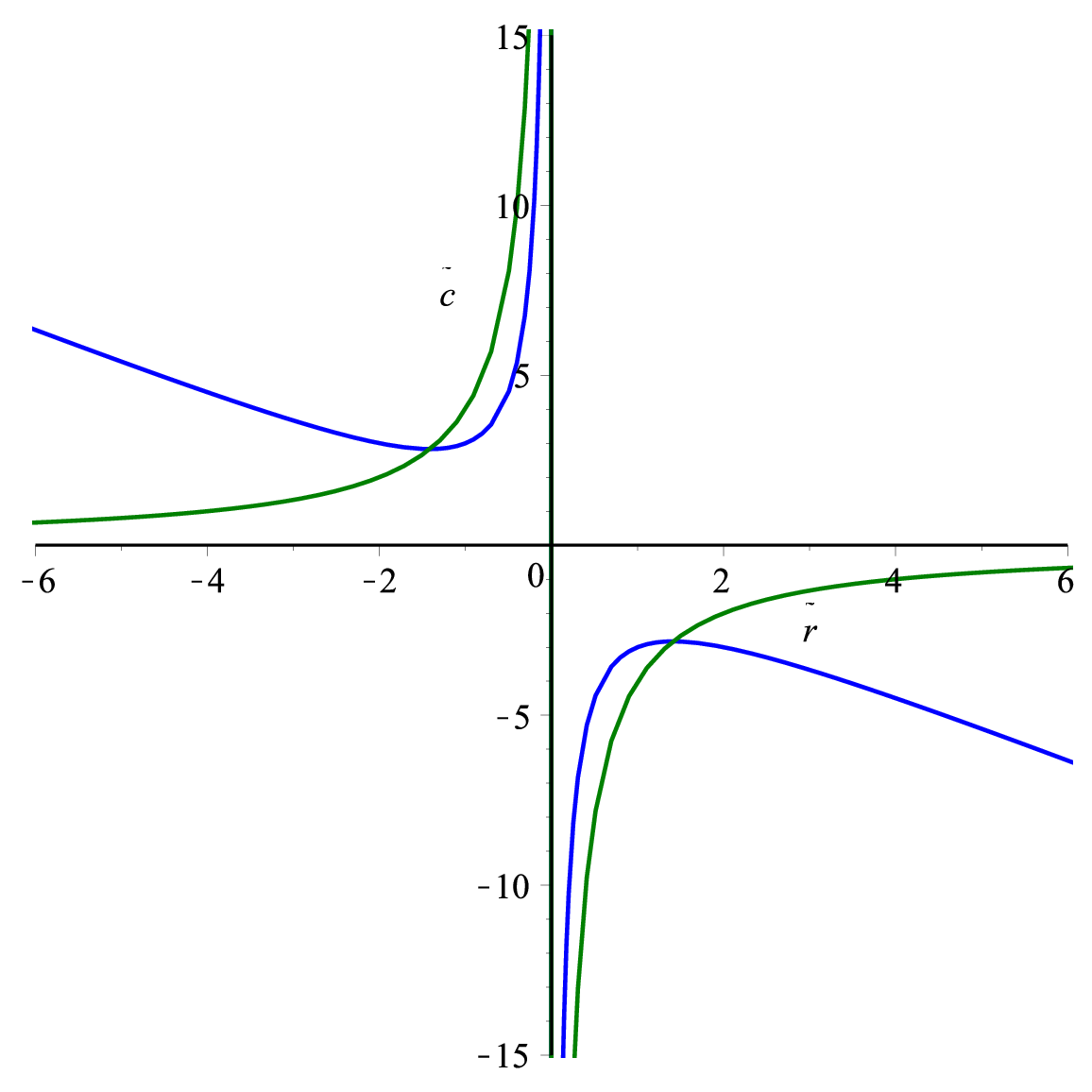}
	}
	\subfigure[ $ d<0 $]{
		\includegraphics[width=0.4\textwidth]{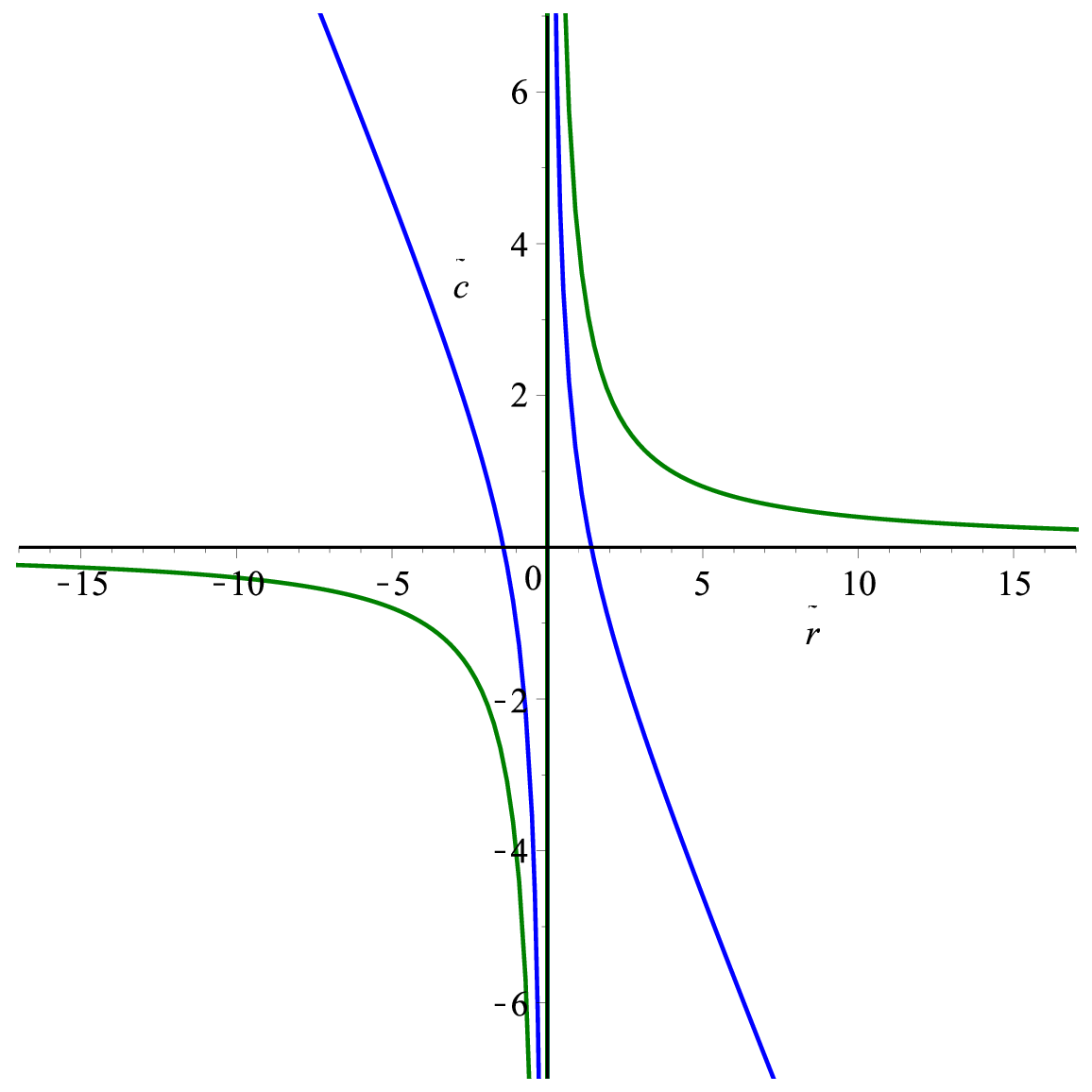}
	}
	\caption{Dimensionless plot of  $ \tilde{c} $ parameter versus the radius $ \tilde{r} $, showing features of the spacetime geometry. (a) for $ d>0 $ and (b) for $ d<0 $. The horizons are in blue, stable light spheres in green for $ \tilde{a_{0}}=\frac{10^{-5}}{3} $ and $ \mathcal{L}=1 $ .}
	\label{cr}
\end{figure}

Plots of effective potentials of photon orbits and the positions of light spheres are shown in Fig.~\ref{FVD}.
Attempts have also been made to show the horizons' locations and the light spheres as a function of the dimensionless parameter $\tilde{c}$, in Fig. \ref{cr}.
Besides, a summary of the results obtained  for the horizons' locations and the positions of stable and unstable light spheres is given in Tab. \ref{tab:HS}. It can be seen from Figs. \ref{FVD}, \ref{cr}, and Tab. \ref{tab:HS} that the values of $\tilde{c}$ and $\tilde{d}$ characterize various spacetime geometries. However some domains such as ($\tilde{c}>0$, $\tilde{d}\geq 0$) and ($\tilde{c}<0$, $\tilde{d}>|c|$) have no horizons, the other domain ($\tilde{c}>0$, $\tilde{d}<0$) has only one event horizon with one stable light sphere and one unstable light sphere, and the domain ($\tilde{c}<0$, $\tilde{d}<0$)has also one event horizon with only one stable light sphere. Other domains have two event horizons (‘interior’ and ‘exterior’ horizons)($\tilde{c}<0$, $0<\tilde{d}<|c|$), with two stable light spheres (‘interior’ and ‘exterior’ light spheres), and one unstable light sphere.

\begin{table}
	\begin{center}
		\begin{tabular}
			{|c|c|c|c|c|c|c|}
			\hline
			\multicolumn{2}{|c|}{Parameters}  & \multicolumn{2}{|c|}{Event Horizons} & \multicolumn{3}{|c|}{Light Spheres}
			\\ \hline
			$ c $& $ d $ & Int-Horizon  & Out-Horizon & Int-Stable & Unstable & Out-Stable
			\\ \hline
			$c<0$ & $ d>|c| $ &  &  &  &  &
			\\ \hline
			$c<0$& $ d=|c| $ & & $ \times $ &  &  & $ \times $  \\ \hline
			$c<0$ & $ d=0 $ &  & $ \times $ &  &  & $ \times $  \\ \hline
			$c<0$& $ 0<d<|c| $ & $ \times $ & $ \times $ & $ \times $ & $ \times $ & $ \times $ \\ \hline
			$c<0$& $d<0$ &  & $ \times $ &  &  & $\times $
			\\ \hline
			$c>0$& $d\geq 0$ &  &  &  &  &
			\\ \hline
			$c>0$& $d<0$ &  & $ \times $ & $ \times $ & $ \times $ &
			\\\hline
		\end{tabular}
		\caption{Possible radial configurations of horizons and light spheres in a SU(2)-colored (A)dS black hole in conformal gravity for $ \tilde{a_{0}}=\frac{10^{-5}}{3} $, $ \mathcal{L}=1 $ and different values of the $ c $ and $ d $ parameters.}\label{tab:HS}
	\end{center}
\end{table}

\clearpage

\section{Analytical solution for motion of  test particles}\label{AS}
One can solve the geodesic equations analytically using elliptic and hyperelliptic
functions. With the dimensionless quantities (\ref{DP}), Eq.~(\ref{rphi}) can be written as
\begin{eqnarray}\label{Rtild}
(\frac{d\tilde{r}}{d\phi})^2 &=&(\epsilon
\mathcal{L}\tilde{a_0})\tilde{r}^6-2\mathcal{L}\epsilon
\tilde{a_0}\tilde{c} \tilde{r}^5
-(-E^2\mathcal{L}+\tilde{a_0}+\tilde{a_0}(\tilde{c}^2+2\tilde{d})\epsilon \mathcal{L})\tilde{r}^4\\
&-&(2\tilde{a_0}\tilde{c}+2\tilde{a_0}\tilde{d}\tilde{c}\epsilon
\mathcal{L})\tilde{r}^3
-(\tilde{a_0}(\tilde{c}^2+2\tilde{d})+\tilde{a_0}\tilde{d}^2\epsilon
\mathcal{L}))\tilde{r}^2-
2\tilde{a}_0\tilde{d}\tilde{c}\tilde{r}-\tilde{a_0}\tilde{d}^2=R(\tilde{r}).\nonumber
\end{eqnarray}
Eq.~(\ref{Rtild}) implies that $ R(\tilde{r})\geq 0$ is a necessary condition for a geodesic existence. Thus, the real and positive zeros of $ R(\tilde{r})$ are considered. With the substitution
\begin{equation}\label{xi}
r=\pm \frac{1}{\xi}+r_{R},
\end{equation}\label{0}
where $r_{R}$ is the zero of polynomial $ R(\tilde{r})$, the positive sign in Eq.~(\ref{xi}) can be chosen for
$\epsilon = 0$, and the resulting equation is
\begin{equation}\label{epsilon0}
(\frac{d\xi}{d\varphi})^2=\sum_{j=0}^3m_j \xi^j ,\quad
m_j=\frac{1}{(4-j)!}\frac{d^{(4-j)}R}{dr^{(4-j)}}(r_{R}),
\end{equation}
and for $\epsilon =1$ we obtain
\begin{equation}\label{epsilon1}
(\frac{\xi d\xi}{d\varphi})^2=\sum_{j=0}^5n_j \xi^j ,\quad
n_j=\frac{{(-1)}^j}{(6-j)!}\frac{d^{(6-j)}R}{dr^{(6-j)}}(r_{R}).
\end{equation}
\subsection*{Null geodesics }
The Eq.~(\ref{epsilon0}) can be solved as
\begin{equation}
r(\varphi)=\frac{m_3}{4\wp(\varphi -\varphi_{in})-\frac{m_2}{3}}+r_{R},
\end{equation}
where $\wp$ is the Weierstrass function \cite{M.Abramowitz,E. T.
Whittaker}, and
\begin{equation}
\varphi_{in}=\varphi_0+\int_{y_0}^\infty
\frac{dz}{\sqrt{4y^3-g_2-g_3}},\quad y_0=\frac{1}{4}(\frac{\pm
m_3}{r_0-r_{R}}+\frac{m_2}{3}),
\end{equation}
with the standard expressions  of $g_2$ and $g_3$ as
\begin{eqnarray}
g_2&=&\frac{1}{16}(\frac{4}{3}m_2^2), \\
g_3&=&\frac{1}{16}(\frac{1}{3}m_1m_2m_3-\frac{2}{27}m_2^3-m_0m_3^2).
\end{eqnarray}

\subsection*{Timelike geodesics}
The analytic solution of Eq.~(\ref{epsilon1}) can be written as
\cite{Soroushfar:2015wqa,Hackmann:2008zz,Enolski:2010if}
\begin{equation}
    u(\varphi) = \left. \frac{\sigma_1 (\boldsymbol{\varphi}_\infty)}{\sigma_2 (\boldsymbol{\varphi}_\infty)} \right|
    _{ \sigma (\boldsymbol{\varphi}_\infty)=0} \, ,
\end{equation}
with
\begin{equation}
    \boldsymbol{\varphi}_\infty =
    \left(
    \begin{array}{c}
      \varphi_2 \\
      \varphi-\varphi_{\rm in}'
    \end{array}
    \right),
\end{equation}
where $ \varphi_{\rm in}'=  \varphi_{\rm in}+\int_{ \varphi_{\rm
in}}^{\infty}\! \frac{\xi \, \mathrm{d} \xi'}{\sqrt{ P_5(\xi')}}$
and the component $ \varphi_2$ appears with the condition $\sigma
(\boldsymbol{\varphi}_\infty)=0$. Also,
\begin{equation}
\sigma(z)=Ce^{zt}kz \theta [g,h](2\omega^{-1}z;\tau),
\end{equation}
where the function $\sigma_i$, which has characteristic $[g,h]$, is
the $i$th place derivative of Kleinian $\sigma$ function, and it is
given by the Riemann $\theta$-function. Besides, the $\tau$, $(2\omega,
2\acute{\omega})$, and $(2\eta, 2\acute{\eta})$ are related
to the symmetric Riemann matrix, period-matrix, and period-matrix of the
second kind, respectively. Also, $\kappa=\eta(2\omega)^{-1}$ is a matrix, and
$2[g,h] = (0, 1)^t+(1,1)^t\tau$ corresponds to the vector of Riemann
constants with the base point at infinity. The constant $C$, can be
given clearly, see e.g. \cite{Bhattacharyya:2012wz}.
This is the analytic solution of the equation of motion of a test
particle in the spacetime of a SU(2)-colored (A)dS black hole in
CG.

\section{Orbits}\label{O}
\begin{figure}[h]
    \centering
    \subfigure[]{
        \includegraphics[width=0.4\textwidth]{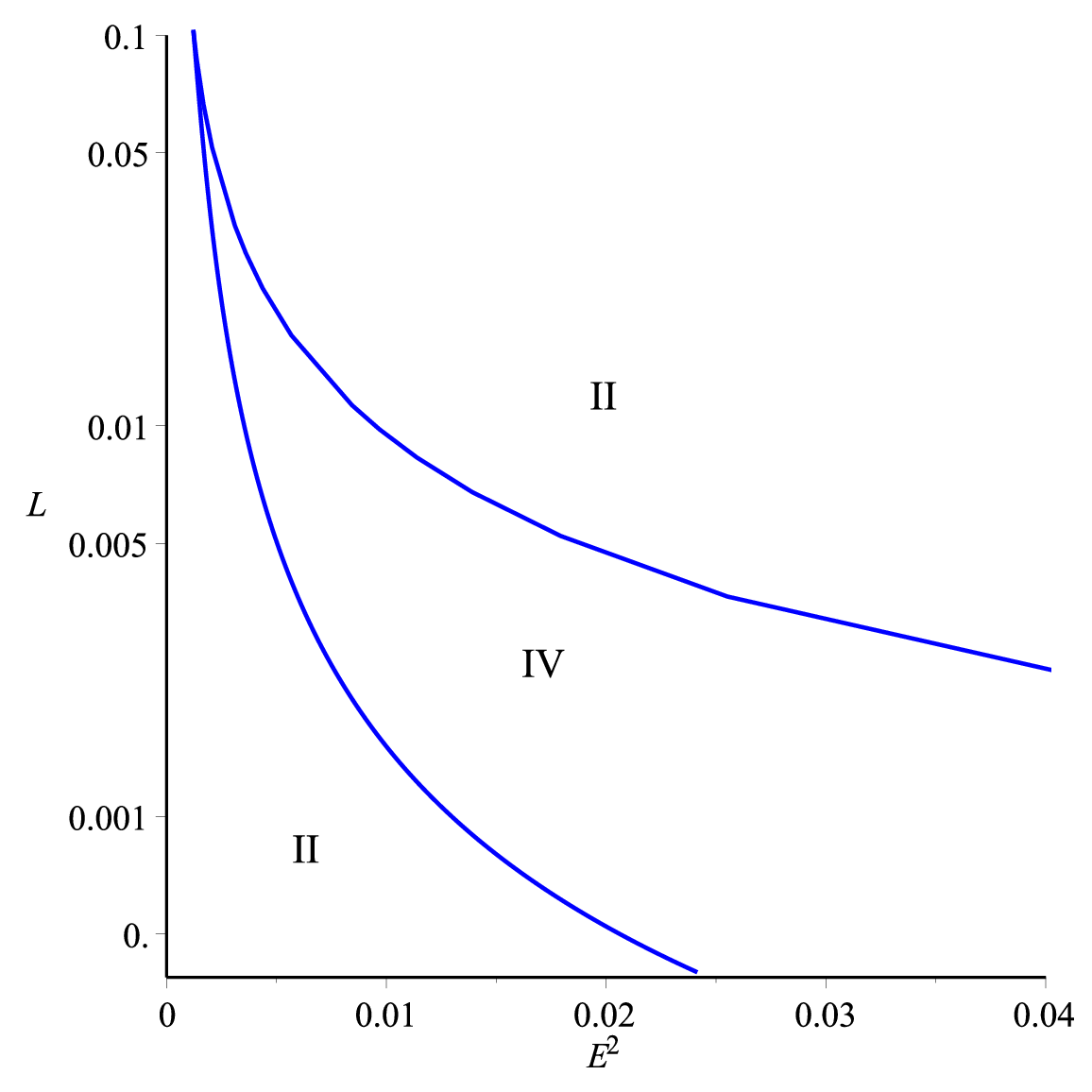}
    }
    \subfigure[]{
        \includegraphics[width=0.4\textwidth]{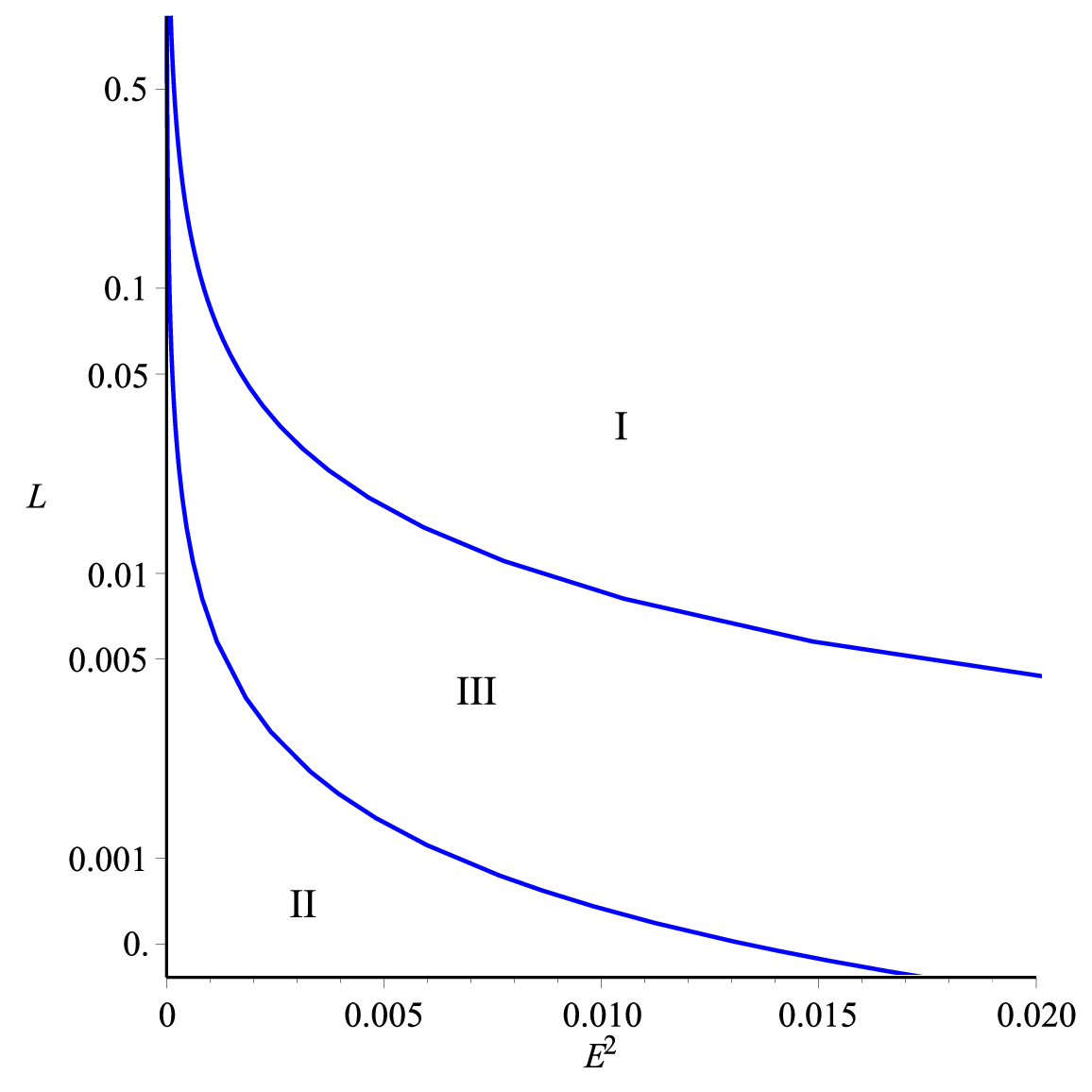}
    }
    \caption{Regions of different types of geodesic motion. (a): for test particles ($ \epsilon=1 $),
        the numbers of positive real zeros in these regions are: II=2, IV=4.
    (b): for light rays ($ \epsilon=0 $),
    the numbers of positive real zeros in these regions are:
    I=1, II=2, and III=3. }
    \label{region}
\end{figure}

With the comparison coefficients of $R(\tilde{r})=0$ and
$\frac{dR(\tilde{r})}{d\tilde{r}}=0$, one can solve the equations
for $E^2$ and $\mathcal{L}$ as
\begin{equation}\label{EL}
\mathcal{L}=-\frac{\tilde{c}\tilde{r}+2\tilde{d}}{(-\tilde{r}^2+\tilde{d})\tilde{r}^2},\quad
E^2=\frac{\tilde{a}(\tilde{c}\tilde{r}+\tilde{r}^2+\tilde{d})^3}{(\tilde{c}\tilde{r}+2\tilde{d})\tilde{r}^2},
\end{equation}
where $c$ and $d$ are defined in Eq.~(\ref{cd}). The plots of $
\mathcal{L}-E^{2} $ diagrams are shown in Fig.~\ref{region}, and
summary of possible orbit types can be found in
Tab.\ref{tab:cyl.orbits}.

\begin{figure}[h]
    \centering
    \subfigure[]{
        \includegraphics[width=0.22\textwidth]{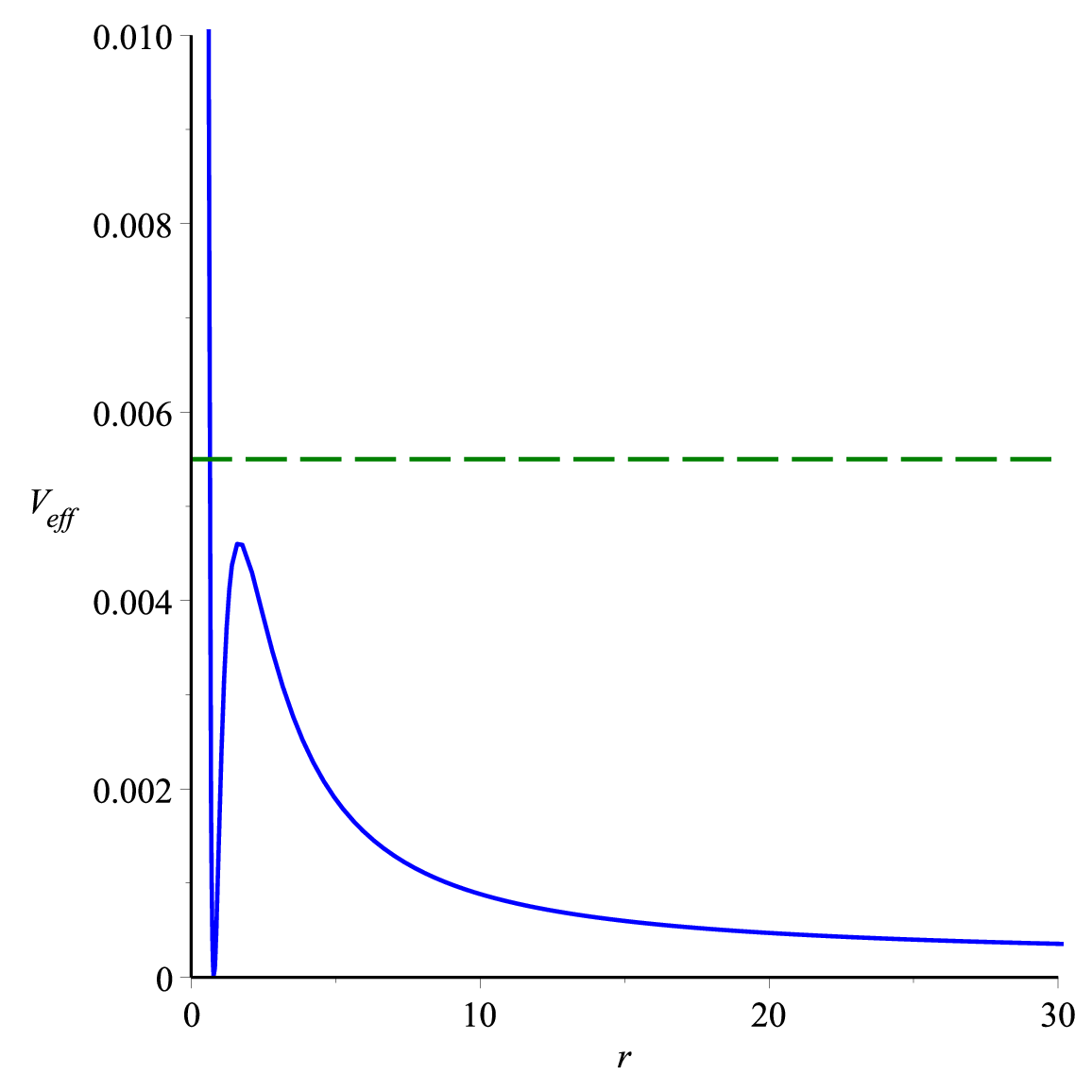}
    }
    \subfigure[]{
        \includegraphics[width=0.22\textwidth]{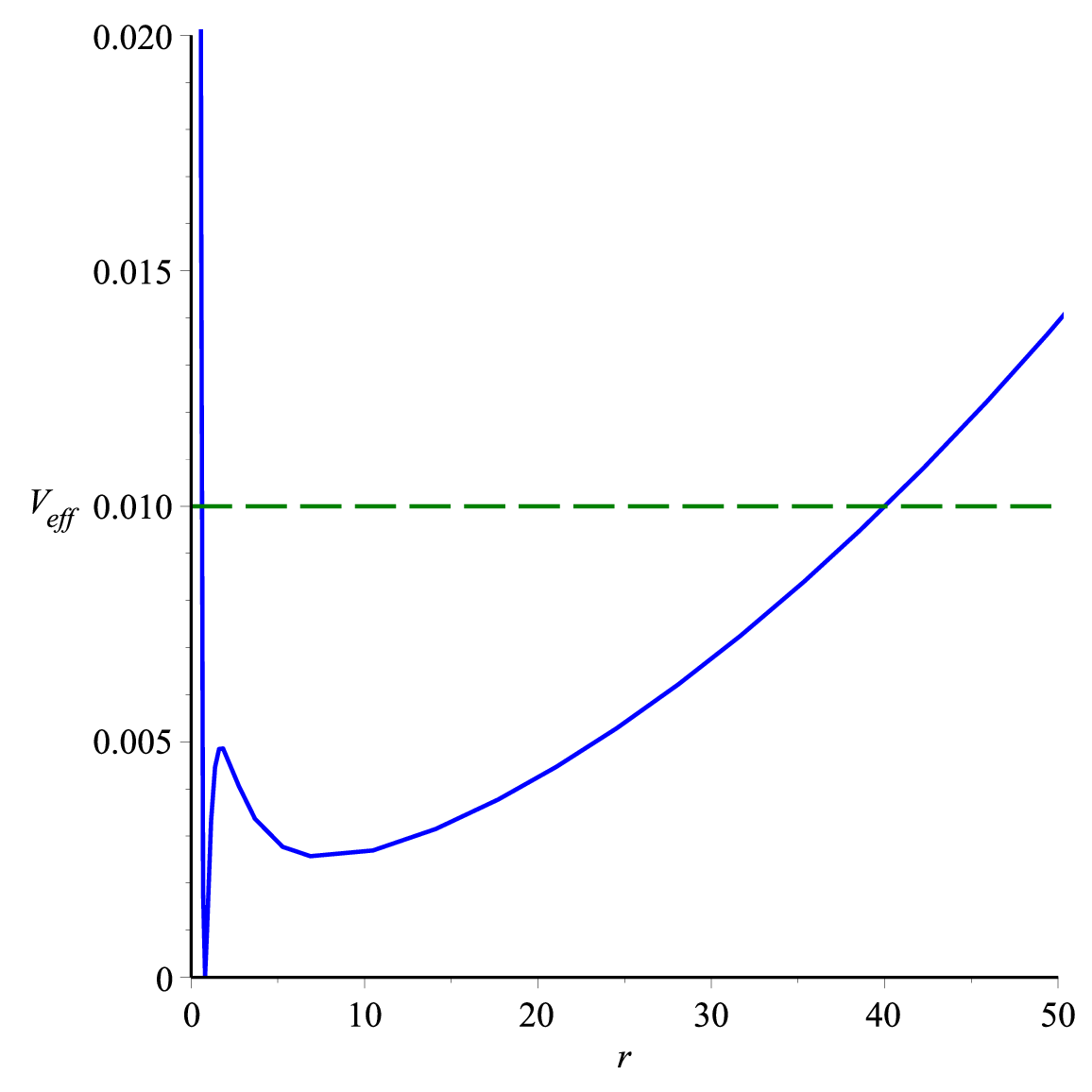}
    }
      \subfigure[]{
       \includegraphics[width=0.22\textwidth]{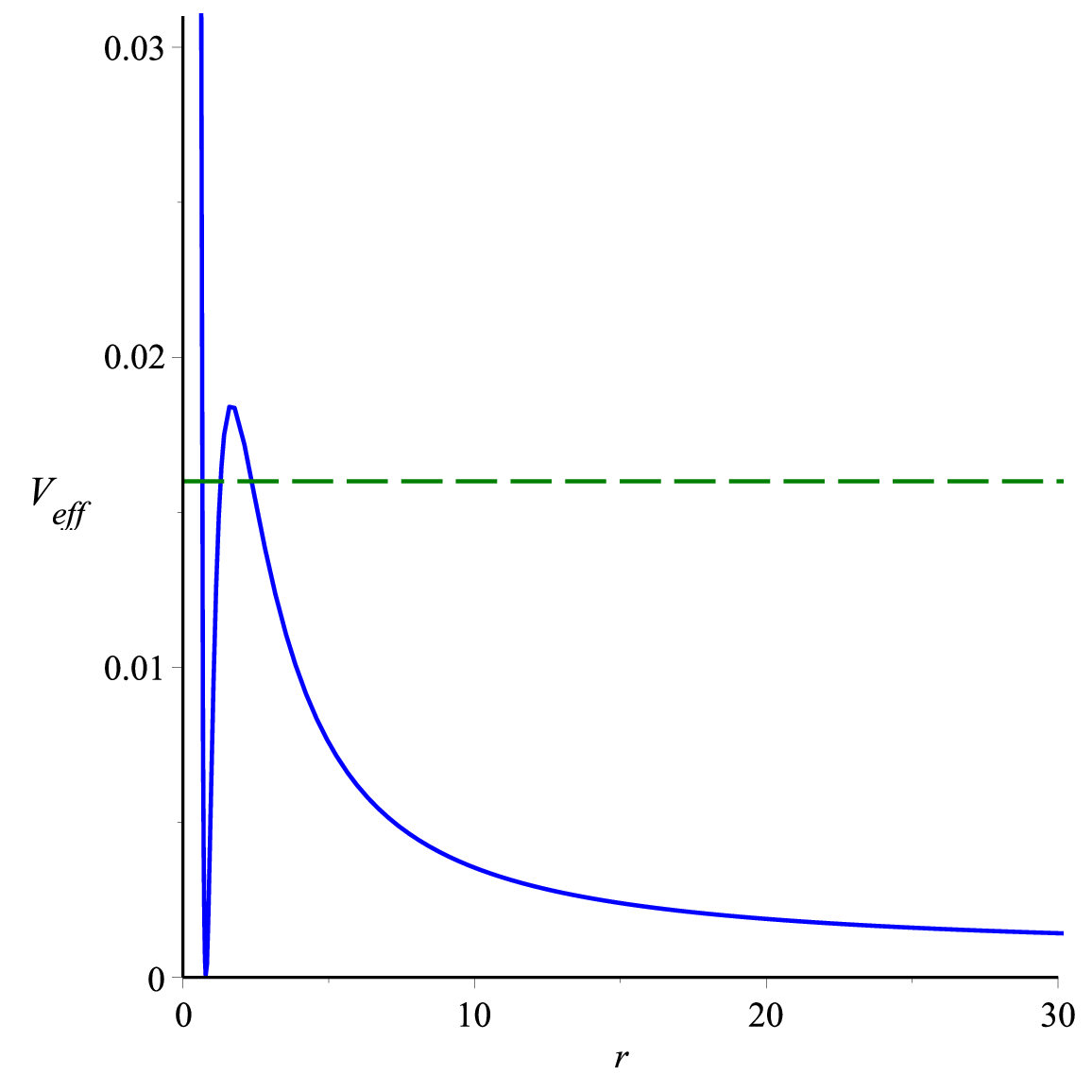}
      }
       \subfigure[]{
       \includegraphics[width=0.22\textwidth]{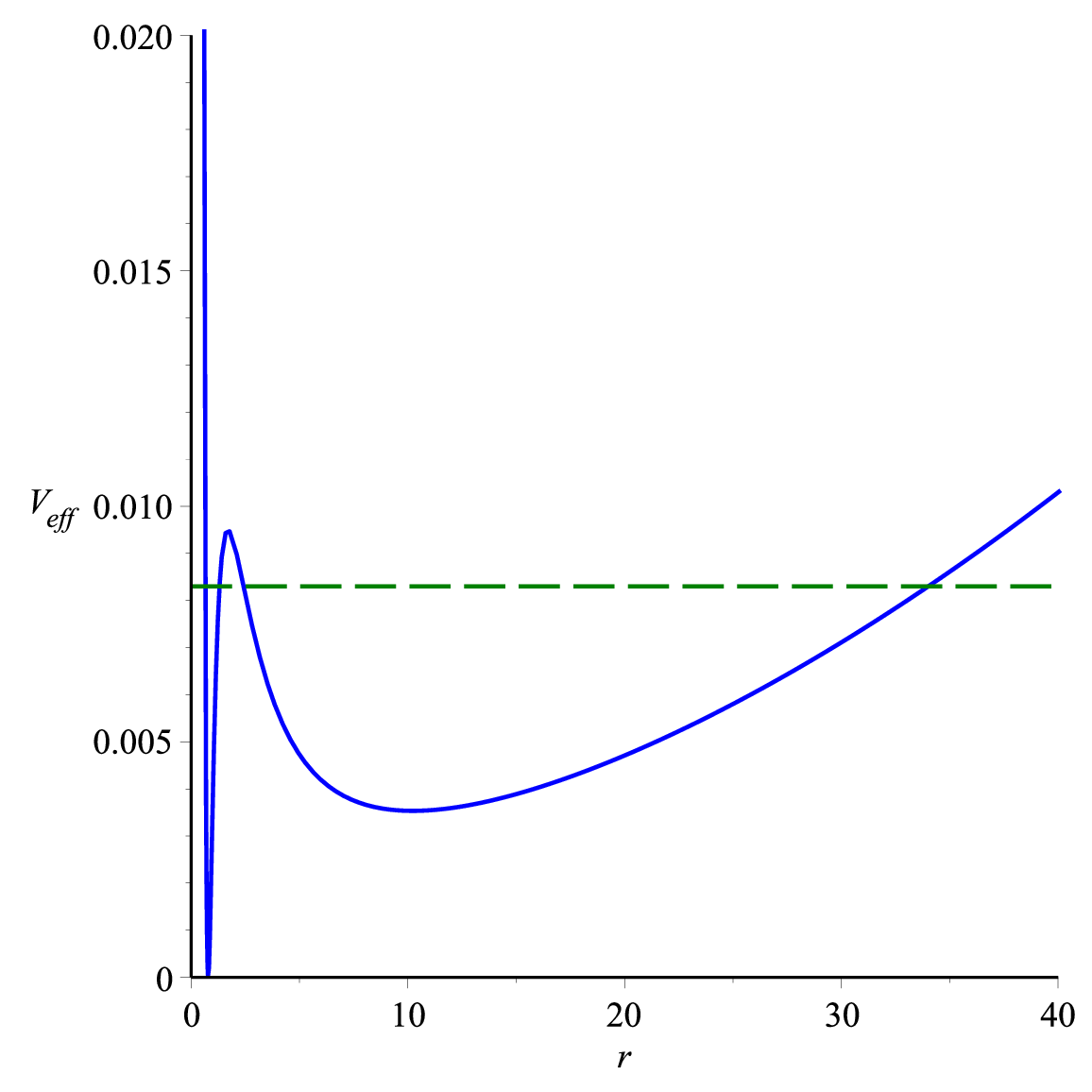}
  }
    \caption{Effective potentials for diffrent regions of geodesic motion. (a), (b), (c) and (d), represent I, II, III, and IV regions of geodesic motion respectively. The horizontal green line represents the squared energy parameter $E^2$.}
    \label{Veff}
\end{figure}

\begin{figure}[h]
    \centering
    \subfigure[]{
        \includegraphics[width=0.3\textwidth]{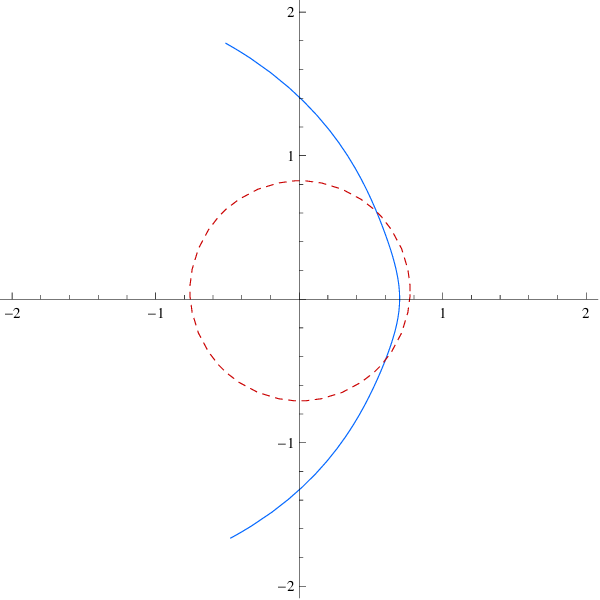}
    }
    \subfigure[]{
        \includegraphics[width=0.3\textwidth]{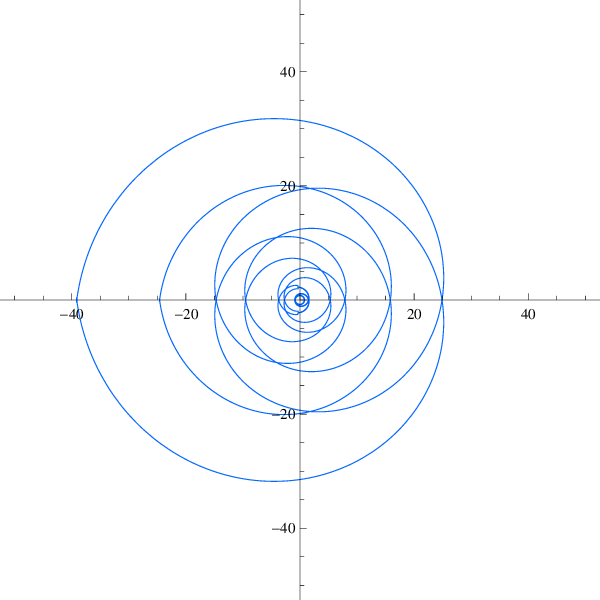}
    }
    \subfigure[]{
        \includegraphics[width=0.3\textwidth]{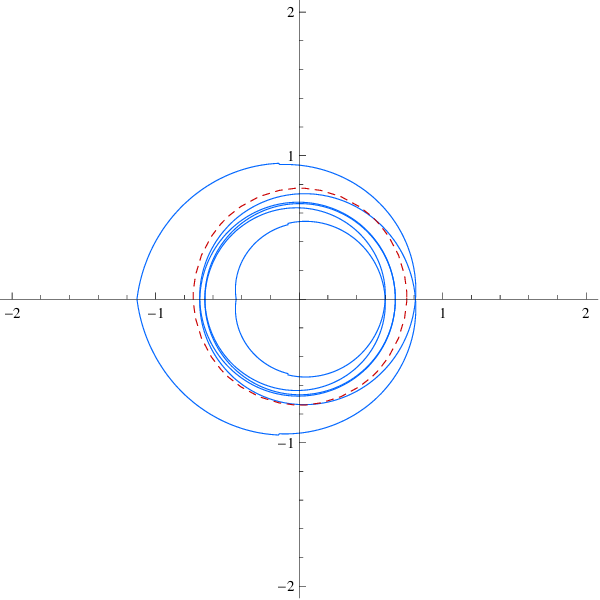}
    }
    \subfigure[]{
        \includegraphics[width=0.3\textwidth]{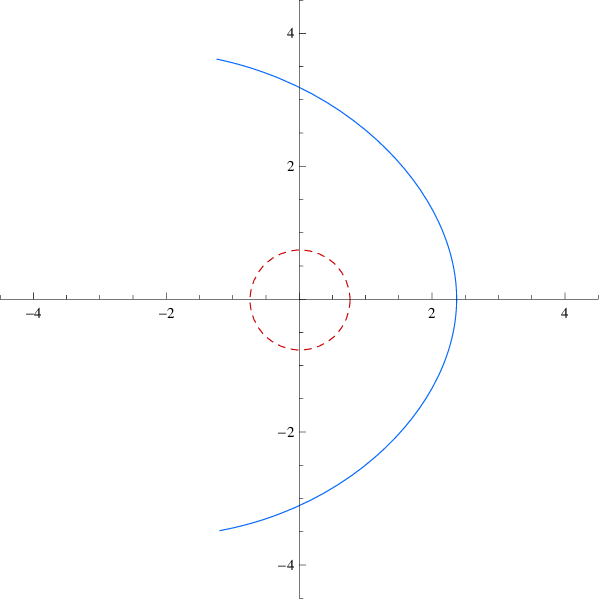}
    }
    \subfigure[]{
        \includegraphics[width=0.3\textwidth]{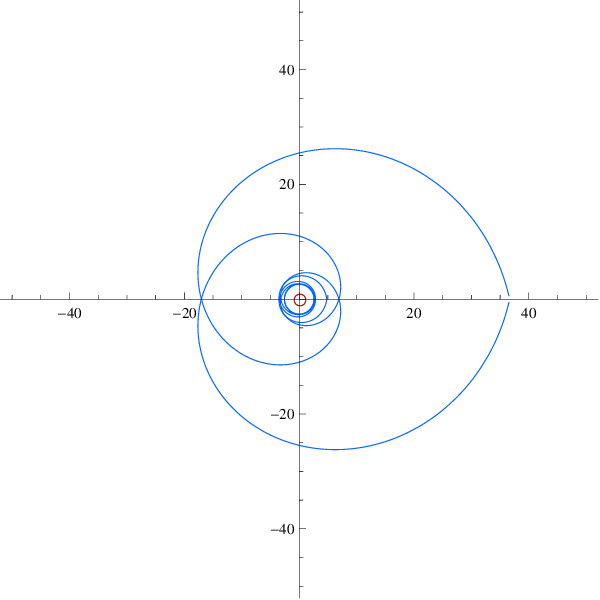}
    }
    \subfigure[]{
        \includegraphics[width=0.3\textwidth]{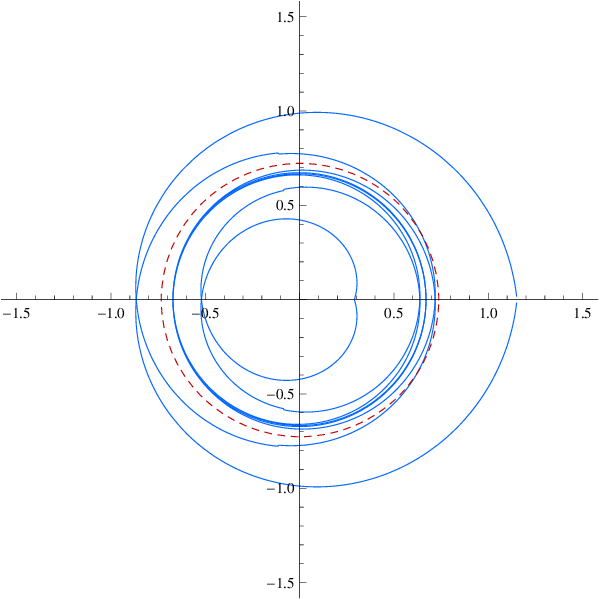}
    }
    \caption{Examples of different orbits in different regions.
    (a): an escape orbit for region I with $\tilde{a_0}=\frac{10^{-5}}{3},\tilde{b_1}=0.9,\tilde{b_0}=-1.5,E^2=0.0055,\mathcal{L}=0.02,\epsilon=0$.
    (b): a bound orbit for region II with $\tilde{a_0}=10^{-5},\tilde{b_1}=-1.5,\tilde{b_0}=0.9,E^2=0.003,\mathcal{L}=0.001,\epsilon=1$.
    (c) and (d): a bound orbit and an escape orbit for region III with $\tilde{a_0}=\frac{10^{-5}}{3},\tilde{b_1}=-1.5,\tilde{b_0}=0.9,E^2=0.016,\mathcal{L}=0.005,\epsilon=0$.
    (e) and (f): two bound orbits for region IV with the parameters $E^2=0.0083, \mathcal{L}=0.01, \varepsilon=1,\tilde{b_1}=-1.5,\tilde{b_0}=0.9,\tilde{a_0}=\frac{10^{-5}}{3}$.}
    \label{Orbit}
\end{figure}

In the following, two types of possible orbits are introduced:
\begin{enumerate}
    \item \textit{Escape Orbit} (EO) with range $\tilde{r} \in [r_1, \infty)$.
    \item \textit{Bound Orbit} (BO) with range $\tilde{r} \in [r_1, r_2]$.
\end{enumerate}
Also, the following classes of regions exists with respect to $r$:
\begin{enumerate}
\item In region I, $R(r)$ has 1 positive real zeros $r_1>0$. The region $[r_1,+\infty)$ correspond to one escape orbit (EO).
\item In region II, $R(r)$ has 2 positive real zeros $r_2>r_1>0$. The region $[r_1,r_2]$ correspond to one bound orbit (BO).
\item In region III, $R(r)$ has 3 positive real zeros $r_3>r_2>r_1>0$. The region $[r_1,r_2]$ correspond to one bound orbit (BO) and the region $[r_3,\infty)$ correspond to one escape orbit (EO).
\item In region IV, $R(r)$ has 4 positive real zeros $r_4>r_3>r_2>r_1>0$. The regions $[r_1, r_2]$ and $[r_3,r_4]$  correspond to two different bound orbits (BO).
\end{enumerate}
As shown in Fig. \ref{region}~ and Tab.~\ref{tab:cyl.orbits}, for
$b_1<0, b_0>1$, there are two regions for timelike geodesics and
three regions for null geodesics. In Fig.~\ref{Orbit}, some of the
possible orbits and light rays are plotted. The parameters
$E^2=0.0055$ and $\mathcal{L}=0.02$, belong to the region I in
Fig.~\ref{region}, corresponding to an escape orbit, (see
Figs. \ref{Veff} (a) and \ref{Orbit} (a)). In region II, which
effective potential is shown in Fig.~\ref{Veff}(b), its orbit is
shown in Fig.~\ref{Orbit} (b). For region III, with three zeroes,
the effective potential is introduced in Fig.~\ref{Veff} (c). The
orbits for this region by considering $E^2=0.016$ and
$\mathcal{L}=0.005$, are indicated in Figs.~\ref{Orbit} (c) and (d).
In region IV, there are four zeroes with two bound orbits; one bound
orbit is related to the two first roots of $R(r)$, and the other one
is related to the two second roots of $R(r)$ (see Figs.~\ref{Veff}
(d) and \ref{Orbit} (e) and (f)). Also, the perihelion shift has
appeared in these two orbits.

\begin{table}[h]
    \begin{center}
        \begin{tabular}{|c|c|c|c|c|}
            \hline
            region & sign of $\beta$, $\gamma$ & positive zeros  & range of $\tilde{r}$ & orbit \\
            \hline\hline
            I & $b_1<0, -1<b_0<1$ & 1 &$|$$--$$\bullet$$\textbf{--------}$$\lVert$$\textbf{-----------------------}$
            & EO
            \\  \hline
            II & $b_1<0, -1<b_0<1$ & 2 &
            $|$$--$$\bullet$$\textbf{--------}$$\lVert$$\textbf{---}$$\bullet$$-------$
            & BO
            \\ \hline
            III  & $b_1<0, -1<b_0<1$ & 3 &
            $|$$--$$\bullet$$\textbf{--------}$$\lVert$$\textbf{---}$$\bullet$$--$$\bullet$$\textbf{------------}$
            & BO, EO
            \\ \hline
            IV & $b_1<0, -1<b_0<1$ & 4 &
            $|$$--$$\bullet$$\textbf{--------}$$\lVert$$\textbf{---}$$\bullet$$--$$\bullet$$\textbf{------}$$\bullet$ $--$
            & 2BO
            \\ \hline
        \end{tabular}
        \caption{Types of orbits in the spacetime of a SU(2)-colored (A)dS black hole. The range of the orbits is represented by thick lines. Dots show the turning points of the orbits. Single vertical line indicates the singularity at $\tilde{r}=0$. The event horizon, is marked by a double vertical line.}
        \label{tab:cyl.orbits}
    \end{center}
\end{table}

\clearpage

\section{Conclusion}\label{C}
This paper derived the geodesics equations of a SU(2)-colored (A)dS black hole in conformal gravity. After reviewing the spacetime and the geodesic equations, we have investigated the spacetime features such as light spheres and horizons. Next, with Weierstrass elliptic and derivatives of Kleinian sigma functions, the analytical solutions for both timelike and lightlike geodesic motion were obtained. Further, a set of orbit types for test particles moving on the geodesic were classified. The analysis realized three regions for null geodesics and two regions for timelike geodesics in the spacetime of a SU(2)-colored (A)dS black hole. The (EO) and (BO) were the possible orbits for null and timelike geodesics.


\bibliographystyle{amsplain}

\end{document}